\documentclass[prd,superscriptaddress,amsfonts,amssymb,amsmath,showpacs,twocolumn]{revtex4-2}
\usepackage{bm}
\usepackage{amsfonts}
\usepackage{latexsym}
\usepackage[latin1]{inputenc}
\usepackage{graphicx}
\usepackage{amsmath}
\usepackage{palatino}
\usepackage{mathpazo}
\usepackage{textcomp}
\usepackage{float}
\usepackage{booktabs}
\usepackage{dcolumn}
\usepackage{booktabs}
\usepackage{multirow}
\usepackage{hyperref}
\hypersetup{colorlinks,citecolor=blue}
\usepackage{amsmath}
\usepackage{xcolor}
\usepackage{orcidlink}
\usepackage[caption=false]{subfig}
\usepackage{commath}
\def\jnl@style{\it}
\def\aaref@jnl#1{{\jnl@style#1}}

\def\aaref@jnl#1{{\jnl@style#1}}

\def\aj{\aaref@jnl{AJ}}                   
\def\apj{\aaref@jnl{ApJ}}                 
\def\apjl{\aaref@jnl{ApJ}}                
\def\apjs{\aaref@jnl{ApJS}}               
\def\apss{\aaref@jnl{Ap\&SS}}             
\def\aap{\aaref@jnl{A\&A}}                
\def\aapr{\aaref@jnl{A\&A~Rev.}}          
\def\aaps{\aaref@jnl{A\&AS}}              
\def\mnras{\aaref@jnl{Mon.~Not.~Roy.~Astron.~Soc.}}             
\def\prd{\aaref@jnl{Phys.~Rev.~D}}        
\def\prc{\aaref@jnl{Phys.~Rev.~C}}  
\def\prl{\aaref@jnl{Phys.~Rev.~Lett.}}    
\def\qjras{\aaref@jnl{QJRAS}}             
\def\skytel{\aaref@jnl{S\&T}}             
\def\ssr{\aaref@jnl{Space~Sci.~Rev.}}     
\def\zap{\aaref@jnl{ZAp}}                 
\def\nat{\aaref@jnl{Nature}}              
\def\aplett{\aaref@jnl{Astrophys.~Lett.}} 
\def\apspr{\aaref@jnl{Astrophys.~Space~Phys.~Res.}} 
\def\physrep{\aaref@jnl{Phys.~Rep.}}      
\def\physscr{\aaref@jnl{Phys.~Scr}}       
\def\commat{\aaref@jnl{Comm.~Math.~Phys.}}              
\def\science{\aaref@jnl{Science}}               
\def\cqg{\aaref@jnl{Classical Quant.~Grav.}}            
\def\jpcs{\aaref@jnl{JPCS}}                                     
\def\ijmpd{\aaref@jnl{Int.~J.~Mod.~Phys.~D}}                    
\def\grg{\aaref@jnl{Gen.~Relat.~Gravit.}}               
\def\rpp{\aaref@jnl{Rep.~Prog.~Phys.}}          
\def\npa{\aaref@jnl{Nucl.~Phys.~A}}        
\def\lrr{\aaref@jnl{Living Rev.~Rel.}}                   
\def\jcap{\aaref@jnl{J.~Cosmology Astropart.~Phys.}}    
\def\rmp{\aaref@jnl{Rev.~Mod.~Phys.}}   
\def\epjc{\aaref@jnl{Eur.~Phys.~J.~C}}

\allowdisplaybreaks[1]

\begin{document}

\color{black}       

\title{Cosmological implications of the constant jerk parameter in $f(Q,T)$ gravity theory}

\author{N. Myrzakulov\orcidlink{0000-0001-8691-9939}}
\email[Email: ]{nmyrzakulov@gmail.com}
\affiliation{L. N. Gumilyov Eurasian National University, Astana 010008,
Kazakhstan.}
\affiliation{Ratbay Myrzakulov Eurasian International Centre for Theoretical
Physics, Astana 010009, Kazakhstan.}

\author{M. Koussour\orcidlink{0000-0002-4188-0572}}%
\email[Email:]{pr.mouhssine@gmail.com}
\affiliation{Quantum Physics and Magnetism Team, LPMC, Faculty of Science Ben
M'sik,\\
Casablanca Hassan II University,
Morocco.}

\author{Alnadhief H. A. Alfedeel\orcidlink{0000-0002-8036-268X}}%
\email[Email:]{aaalnadhief@imamu.edu.sa}
\affiliation{Department of Mathematics and Statistics, Imam Mohammad Ibn Saud Islamic University (IMSIU),\\
Riyadh 13318, Saudi Arabia.}
\affiliation{Department of Physics, Faculty of Science, University of Khartoum, P.O. Box 321, Khartoum 11115, Sudan}
\affiliation{Centre for Space Research, North-West University, Potchefstroom 2520, South Africa}

\author{ H. M. Elkhair\orcidlink{0000-0000-0000-0000}}%
\email[Email:]{eiabdalla@imamu.edu.sa}
\affiliation{Deanship of Scientific Research, Imam Mohammad Ibn Saud Islamic University (IMSIU), \\ P. O Bx 5701, 11432, Riyadh, Saudi Arabia}

\date{\today}

\begin{abstract}
This study delves into modified gravity theories that are equivalent to General Relativity but involve the torsion or non-metricity scalar instead of the curvature scalar. Specifically, we focus on $f(Q,T)$ gravity, which entails an arbitrary function of the non-metricity scalar $Q$ non-minimally coupled to the trace of the stress-energy tensor $T$. We investigate the functional form $f(Q,T)=f(Q)+f(T)$, where $f(Q) =Q+\alpha\,Q^2$ represents the Starobinsky model in $f(Q)$ gravity and $f(T)=2\,\gamma\,T$, with $\alpha$ and $\gamma$ are constants. To obtain solutions for the Friedmann equations, we introduce the concept of a constant jerk and employ its definition to trace the evolution of other kinematic variables, including the deceleration parameter, energy density, EoS parameter, and various energy conditions. These analyses serve to validate the proposed model. We constrain our constant jerk model using the most available Pantheon set of data, the Hubble set of data, and the BAO set of data. Further, we employ $Om(z)$ diagnostic as a means to differentiate between different theories of dark energy. Throughout our examination, all cosmological parameters under scrutiny consistently indicate an accelerating Universe.
\end{abstract}

\maketitle

\section{Introduction}

\label{sec1}

Some recent Type Ia supernovae (SNe) data \cite{Riess, Perlmutter} and
Planck Collaboration \cite{Planck2020} conclusions have altered the cosmic
image of the Universe. The fact that the expansion of the Universe presently
accelerating is a revolutionary indicator of these observations. To
understand this Cosmic acceleration, several ideas have been presented in
the literature. The reason for this acceleration is thought to be the Dark
Energy (DE), which cannot be described by the baryonic matter distribution.
This DE is currently known to account for around $70\%$ of the overall
energy distribution in the Universe. This dark energy is commonly described
in $\Lambda $ cold-dark-matter ($\Lambda$CDM) cosmology by introducing the
cosmological constant $\Lambda$ to the Einstein's Field Equations of the
General Relativity (GR). Moreover, such a cosmic scenario is fraught with
cosmological issues \cite{weinberg/1989}, prompting the development of
alternative models. The model that acts similarly to the $\Lambda$CDM model
is achieved in these alternative cosmologies without the use of the
cosmological constant. The major reason for this is the necessity for a
cosmological model that can provide the findings of the observable Universe
while also avoiding the issues introduced by $\Lambda$CDM. For instance,
taking the matter-energy content of the Universe as the scalar field for
exotic matter in the Einstein's Field Equations that may provide sufficient
negative pressure to accelerate the expansion of the Universe is one of the
implications of such a necessity \cite{Kamenshchik, Sahni, Li, Cai,
Padmanabhan, Caldwell, Nojiri}.

Modified gravity theories (MGT) that suit this objective are of particular
interest in contemporary cosmological research, such as $f(R)$
theory \cite{fR1,fR2}, $f(T)$ theory \cite{fT1,fT2}, and $f(R,T)$ theory 
\cite{fRT}. One of the most widely accepted MGT called $f(Q)$ gravity
introduced by Jimenez et al. \cite{Jimenez/2018}. In this case, $Q$
geometrically represents the variation of a vector's length in parallel
transport. In this paper, we will look at the $f(Q,T)$ gravity theory, which
is an extension of the newly introduced $f(Q)$ theory of gravity \cite{Yixin/2019}. Here, the gravitational Lagrangian is represented by an
arbitrary function of the non-metricity scalar $Q$, and the energy-momentum
tensor trace $T$, the dependence of which can be caused by exotic imperfect
fluids or quantum phenomena \cite{fRT}. The teleparallel extension of GR
gives rise to $f(Q,T)$ gravity. Curvature and non-metricity disappear in the
teleparallel representation, and the metric tensor $g_{\mu\,\nu}$ is
replaced by a collection of tetrad vectors $e_{\mu }^{i}$. Nester and Yo 
\cite{Nester} proposed an alternative equivalent representation called
symmetric teleparallel gravity, in which the geometric variable is
represented by $Q$ and finally structured as $f(Q)$ gravity \cite%
{Jimenez/2018}. Furthermore, the $f(Q,T)$ gravity has been structured using
a non-minimal coupling between non-metricity and the trace of
energy-momentum tensor \cite{Yixin/2019}. By varying the gravitational action with regard to both metric and connections, the theory's field equations were derived. By assuming a simple functional form of $f(Q,T)$, they were able to get the cosmic evolution equations for a flat,
homogeneous, isotropic geometry $f(Q,T)$. Subsequently, $f(Q, T)$ gravity has been extensively examined in various cosmological and astrophysical contexts, including investigations of Cosmic acceleration \cite{Zia,Godani}, Baryogenesis \cite{Bhattacharjee}, Cosmological inflation \cite{Inflation}, Static spherically symmetric wormholes \cite{Tayde}, Cosmological perturbations \cite{Najera}, Energy conditions \cite{Arora2}, Observational constraints on $f(Q, T)$ gravity models \cite{Arora3}, and Matter bounce cosmology \cite{Agrawal}. While many studies have focused on examining this theory with observational datasets, relatively few have explored its applications to kinematical variables such as the constant deceleration or jerk parameters \cite{jerkf}.

In this paper, we analyse the Starobinsky cosmological $f(Q,T)$ model with a spatially flat homogeneous and isotropic geometry using $f(Q,T)=Q+\alpha\,Q^2+2\,\gamma\,T$ and this backed by $f(R,T)$ gravity form $%
f(R,T)=R+\alpha\, R^2+2\,\gamma\,T$ where the presence of the square term of $R$ confirms the existence of DE and dark matter. Then,
we introduce the basic kinematical variables such as the Hubble parameter $%
H(t)$, the deceleration parameter $q(t)$, and the jerk parameter $j(t)$, are
just the first, second, and third order time derivatives of the scale factor 
$a(t)$, respectively. All of the derivatives are fractional, and $q(t)$ and $%
j(t)$ are dimensionless \cite{Blandford}. Rapetti et al. \cite{Rapetti} conducted a thorough analysis of the jerk parameter as a means of developing a model to investigate the Universe's expansion history. Zhai et al. \cite{Zhai} recently constrained four jerk models using various parametrizations of $j(z)$ ($j=$ $\Lambda$CDM value + departure) as a function of redshift $z$ by establishing the $\Lambda$CDM model as the fiducial model and employing Type Ia Supernova with 580 data points and observational Hubble parameter data with 21 data points. Here, we consider a constant jerk and
use the definition of jerk to determine the evolution of the other kinematics variables. The values of the different kinematics variables and
model parameters, which are expressed as the constant of integration and the
value of $j$, are then approximated using existing observational sets of
data. This approach is also known as the model-independent way study of
cosmological models (or the cosmological parametrization) \cite{Pacif1,
Pacif2}, and it generally assumes parametrizations of any kinematic
variables such as $H\left( t\right) $, $q\left( t\right) $, $j\left(
t\right) $ and $\omega \left( t\right) $ (EoS parameter) and provides the
required supplementary equation to solve the system of field equations
completely \cite{Koussour1, Del, Mukherjee}. As data sets get larger, researchers study DE parametrization. The most of research has focused on
observable evidence from SNe, Cosmic Microwave Background (CMB), and the Baryon Acoustic Oscillation (BAO), all of which have been shown to be useful in
constraining cosmological models. The Hubble parameter $H(z)$ set of data
reveals the complicated structure of the expansion of the Universe. The ages
of the most massive and slowly developing galaxies provide direct
measurements of the $H(z)$ at different red shifts $z$, culminating in the
construction of a new type of standard cosmological probe \cite{jim}. Nonetheless, our study is more general and, in some respects, distinct from other similar works \cite{Mukherjee1,Mukherjee2}. To begin, a fundamental difference between our study and that of Zhai et al. \cite{Zhai}, is that we do not assume an a priori flat $\Lambda$CDM model for the current Universe, but rather enable our model to act more broadly. Furthermore, the observational data can be applied to fix the current value of the jerk parameter. Second, in this paper, we investigate the evolution of the jerk parameter in the framework of $f(Q,T)$ gravity. Finally, we present the updated Hubble $H(z)$ set of data, which contain 31 data points using the differential age approach \cite{Hubble,Sharov}, the recently published Pantheon set of data, which contain 1048 data points across the red shift
range $z\in \lbrack 0.01,2.3]$ \cite{Scolnic}, and the BAO set of data, which contain six data points \cite{BAO1, BAO2, BAO6}. The $H(z)$, BAO, and SNe are
used in our analysis to constrain the cosmological model. 

The paper is structured as follows: In Sec. \ref{sec2}, we provide an
outline of $f(Q,T)$ gravity. In Sec. \ref{sec3}, we propose the cosmological
model used in the paper, along with certain model parameters, and we derive
several physical parameters. In Sec. \ref{sec4}, we use $H(z)$, SNe, and BAO
data sets to constrain the model parameters. Further, in Sec. \ref{sec5},
we analyze the behavior of EoS parameter and different energy condition to
validate the proposed model. In Sec. \ref{sec6}, we examine the behavior of
$Om(z)$ diagnostic on values constrained by observational data in order to
distinguish between DE models. Finally, in Sec. \ref{sec7}, we describe our
findings.

\section{$f(Q,T)$ gravity theory}

\label{sec2}

Here, we briefly discuss the modified $f(Q,T)$ gravity using the method in 
\cite{Yixin/2019}. The metric tensor $g_{\mu \nu }$ can be considered as an
extension of the gravitational potential and is mainly used to establish
fundamental concepts such as volumes, distances, and angles. In contrast,
the affine connection $\Gamma ^{\mu }{}_{\alpha \beta }$ is responsible for
parallel transport and covariant derivatives. A fundamental principle in
differential geometry states that the general affine connection can be
separated into three distinct components,%
\begin{equation}
\widetilde{\Gamma }^{\gamma }{}_{\mu \nu }=\Gamma ^{\gamma }{}_{\mu \nu
}+C^{\gamma }{}_{\mu \nu }+L^{\gamma }{}_{\mu \nu }\,.
\end{equation}

In this case, $\Gamma ^{\gamma }{}_{\mu \nu }\equiv \frac{1}{2}g^{\gamma
\beta }\left( \partial _{\mu }g_{\beta \nu }+\partial _{\nu }g_{\beta \mu
}-\partial _{\beta }g_{\mu \nu }\right) $ represents the Levi-Civita
connection of the metric tensor $g_{\mu \nu }$, $C^{\gamma }{}_{\mu \nu
}\equiv \frac{1}{2}T^{\gamma }{}_{\mu \nu }+T_{(\mu }{}^{\gamma }{}_{\nu )}$
represents the contortion tensor, where the torsion tensor is defined as $%
T^{\gamma }{}_{\mu \nu }\equiv 2\Gamma ^{\gamma }{}_{[\mu \nu ]}$, and the
disformation tensor $L^{\lambda }{}_{\mu \nu }$ is represented by 
\begin{equation}
L^{\gamma }{}_{\mu \nu }\equiv \frac{1}{2}g^{\gamma \beta }\left( Q_{\nu \mu
\beta }+Q_{\mu \nu \beta }-Q_{\gamma \mu \nu }\right) \,.
\end{equation}

The non-metricity tensor $Q_{\gamma \mu \nu }$ is defined as the negative of
the covariant derivative of the metric tensor with respect to the
Weyl--Cartan connection $\widetilde{\Gamma }^{\gamma }{}_{\mu \nu }$, i.e. $%
Q_{\gamma \mu \nu }=-\nabla _{\gamma }g_{\mu \nu }$. This tensor can be
derived as,
\begin{equation}
Q_{\gamma \mu \nu }=-\partial _{\gamma }g_{\mu \nu }+g_{\nu \sigma }%
\widetilde{\Gamma }{^{\sigma }}_{\mu \gamma }+g_{\sigma \mu }\widetilde{%
\Gamma }{^{\sigma }}_{\nu \gamma },  \label{4}
\end{equation}%
and the trace of the non-metricity tensor being provided as, 
\begin{equation}
Q_{\beta }=g^{\mu \nu }Q_{\beta \mu \nu },\qquad \widetilde{Q}_{\beta
}=g^{\mu \nu }Q_{\mu \beta \nu }.  \label{5}
\end{equation}

A super-potential or the non-metricity conjugate can also be defined as, 
\begin{eqnarray}
\hspace{-0.5cm} &&P_{\ \ \mu \nu }^{\beta }\equiv \frac{1}{4}\bigg[-Q_{\ \
\mu \nu }^{\beta }+2Q_{\left( \mu \ \ \ \nu \right) }^{\ \ \ \beta
}+Q^{\beta }g_{\mu \nu }-\widetilde{Q}^{\beta }g_{\mu \nu }  \notag \\
\hspace{-0.5cm} &&-\delta _{\ \ (\mu }^{\beta }Q_{\nu )}\bigg]=-\frac{1}{2}%
L_{\ \ \mu \nu }^{\beta }+\frac{1}{4}\left( Q^{\beta }-\widetilde{Q}^{\beta
}\right) g_{\mu \nu }-\frac{1}{4}\delta _{\ \ (\mu }^{\beta }Q_{\nu )}.\quad
\quad   \label{6}
\end{eqnarray}%
expressing the scalar of non-metricity as \cite{Jimenez/2018}, 
\begin{eqnarray}
&&Q=-Q_{\beta \mu \nu }P^{\beta \mu \nu }=-\frac{1}{4}\big(-Q^{\beta \nu
\rho }Q_{\beta \nu \rho }+2Q^{\beta \nu \rho }Q_{\rho \beta \nu }  \notag \\
&&-2Q^{\rho }\tilde{Q}_{\rho }+Q^{\rho }Q_{\rho }\big).  \label{7}
\end{eqnarray}

Symmetric teleparallel gravity is a geometric explanation of gravity that is
entirely equivalent to GR. This equivalence can be demonstrated simply in
the coincident gauge by setting $\widetilde{\Gamma }^{\gamma }{}_{\mu \nu }=0
$. By enforcing the symmetric condition on the connection, the torsion
tensor becomes zero ($T^{\gamma }{}_{\mu \nu }=0$), and the Levi-Civita
connection can be formulated in terms of the disformation tensor as $\Gamma
^{\gamma }{}_{\mu \nu }=-L^{\gamma }{}_{\mu \nu }$. For the $f(Q,T)$ theory,
the action is defined by, 
\begin{equation}
S=\int \sqrt{-g}\left( \frac{1}{16\pi }f(Q,T)+L_{m}\right) d^{4}x,  \label{1}
\end{equation}%
where $f(Q,T)$ is any arbitrary function of $Q$ and $T$. While $L_{m}$ is
the typical matter Lagrangian, $Q$ is the non-metricity scalar, and $T$ is
the trace of energy momentum tensor $T_{\mu \nu }$. The energy-momentum
tensor $T_{\mu \nu }$ is written as, 
\begin{equation}
T_{\mu \nu }=-\frac{2}{\sqrt{-g}}\dfrac{\delta (\sqrt{-g}L_{m})}{\delta
\,g^{\mu \nu }}.
\end{equation}

In addition, the variation of energy-momentum tensor with respect to the
metric tensor is, 
\begin{equation}
\frac{\delta \,g^{\,\mu \nu }\,T_{\,\mu \nu }}{\delta \,g^{\,\alpha \,\beta }%
}=T_{\,\alpha \beta }+\Theta _{\,\alpha \,\beta }.  \label{10}
\end{equation}%
And%
\begin{equation}
\Theta _{\mu \nu }=g^{\alpha \beta }\frac{\delta T_{\alpha \beta }}{\delta
g^{\mu \nu }}.  \label{9}
\end{equation}

As a result, by equating the variation of action \eqref{1} with regard to
the metric tensor to zero, we obtain the following field equations: 
\begin{multline}
-\frac{2}{\sqrt{-g}}\nabla _{\beta }(f_{Q}\sqrt{-g}P_{\,\,\,\,\mu \nu
}^{\beta })-\frac{1}{2}fg_{\mu \nu }+f_{T}(T_{\mu \nu }+\Theta _{\mu \nu })
\label{11} \\
-f_{Q}(P_{\mu \beta \alpha }Q_{\nu }^{\,\,\,\beta \alpha }-2Q_{\,\,\,\mu
}^{\beta \alpha }P_{\beta \alpha \nu })=8\pi T_{\mu \nu }.
\end{multline}%
where $f_{Q}=\dfrac{df}{dQ}$, $f_{T}=\dfrac{df}{dT}$ and $T_{\mu \nu }$ is
the energy-momentum tensor for the fluid of the ideal type, as described
below.

Furthermore, it is worth mentioning that the divergence of the
matter-energy-momentum tensor in the $f(Q,T)$ theory can be expressed as,
\begin{eqnarray}
\hspace{-0.5cm} &&\mathcal{D}_{\mu }T_{\ \ \nu }^{\mu }=\frac{1}{f_{T}-8\pi }%
\Bigg[-\mathcal{D}_{\mu }\left( f_{T}\Theta _{\ \ \nu }^{\mu }\right) -\frac{%
16\pi }{\sqrt{-g}}\nabla _{\alpha }\nabla _{\mu }H_{\nu }^{\ \ \alpha \mu } 
\notag \\
\hspace{-0.5cm} &&+8\pi \nabla _{\mu }\bigg(\frac{1}{\sqrt{-g}}\nabla
_{\alpha }H_{\nu }^{\ \ \alpha \mu }\bigg)-2\nabla _{\mu }A_{\ \ \nu }^{\mu
}+\frac{1}{2}f_{T}\partial _{\nu }T\Bigg],
\end{eqnarray}%
where $H_{\gamma }^{\ \ \mu \nu }$ is the hyper-momentum tensor density
defined as,%
\begin{equation}
H_{\gamma }^{\ \ \mu \nu }\equiv \frac{\sqrt{-g}}{16\pi }f_{T}\frac{\delta T%
}{\delta \widetilde{\Gamma }_{\ \ \mu \nu }^{\gamma }}+\frac{\delta \sqrt{-g}%
\mathcal{L}_{M}}{\delta \widetilde{\Gamma }_{\ \ \mu \nu }^{\gamma }}.
\end{equation}

Thus, the above equation illustrates that in the $f(Q,T)$ gravity theory,
the matter-energy-momentum tensor is not conserved, i.e. $\hspace{-0.1cm}%
\mathcal{D}_{\mu }T_{\ \ \nu }^{\mu }\neq 0$. This non-conservation can be
interpreted as an additional force acting on massive test particles,
resulting in non-geodesic motion. It also indicates the amount of energy
that either enters or exits a specific volume of a physical system.
Furthermore, the non-zero right-hand side of the energy-momentum tensor
implies the presence of transfer processes or particle production in the
system. It is worth noting that the energy-momentum tensor becomes conserved
if $f_{T}$ terms are absent in the aforementioned equation \cite{Yixin/2019}.

Now, suppose the Universe can be represented by the homogeneous, isotropic,
and spatially flat FLRW metric, 
\begin{equation}
ds^{2}=-dt^{2}+a^{2}(t)\left[ dx^{2}+dy^{2}+dz^{2}\right] ,  \label{12}
\end{equation}%
where $a(t)$ is the scale factor of the Universe used to estimate the rate
of cosmic expansion at a time $t$. Further, it is presumed that the known
Universe matter is made up of a perfect fluid, for which the energy-momentum
tensor, $T_{\,\,\,\nu }^{\mu }=diag(-\rho ,p,p,p)$ with its trace $%
T=-\rho+3\,p$. Moreover, the non-metricity scalar $Q$ for this type of
metric is derived and given as $Q=6H^{2}$, where $H$ is the Hubble parameter.

Using the metric (\ref{12}) and the field equation \eqref{11}, the
generalized Friedmann equations are obtained as, 
\begin{equation}
8\pi \rho =\frac{f}{2}-6FH^{2}-\frac{2\widetilde{G}}{1+\widetilde{G}}(\dot{F}%
H+F\dot{H}),  \label{13}
\end{equation}%
\begin{equation}
8\pi p=-\frac{f}{2}+6FH^{2}+2(\dot{F}H+F\dot{H}),  \label{14}
\end{equation}%
where, the dot ($\cdot $) denotes a derivative with respect to time, while
the symbols $F=f_{Q}$, and $8\pi \widetilde{G}=f_{T}$, respectively, signify
differentiation with respect to $Q$, and $T$.

Using the two Eqs. \eqref{13} and \eqref{14} mentioned above , we can
construct the equations similar to the form of standard GR, 
\begin{equation}
3H^{2}=8\pi \rho _{eff}=\frac{f}{4F}-\frac{4\pi }{F}\left[ (1+\widetilde{G}%
)\rho +\widetilde{G}p\right] ,  \label{15}
\end{equation}%
and 
\begin{eqnarray}
2\dot{H}+3H^{2} &=&-8\pi p_{eff}=\frac{f}{4F}-\frac{2\dot{F}H}{F}+  \notag \\
&&\frac{4\pi }{F}\left[ \left( 1+\tilde{G}\right) \rho +\left( 2+\tilde{G}%
\right) p\right] .  \label{16}
\end{eqnarray}%
where the terms $\rho _{eff}$, and $p_{eff}$ refer to the effective pressure
and density, respectively.

\section{Starobinsky cosmological $f(Q,T)$ model}

\label{sec3}

In the $f(R)$ gravity, Starobinsky suggested a reputable and trustworthy
functional form as $f(R)=R+\alpha\,R^2$, this has $\alpha$ as a constant and
is called the Starobinsky model \cite{Starobinsky1,Starobinsky2}. In the
gravitational component of the Einstein-Hilbert action, it predicts that a
quadratic correction of the Ricci scalar will be added. In the literature,
Starobinsky model has been extensively used to cosmological and
astrophysical applications. A cosmological model derived from the previous
function, according to Starobinsky, can pass cosmological observational
tests \cite{Starobinsky2}. Starobinsky model is also quite significant from
an astrophysical perspective. By taking into account axially symmetric
dissipative dust under geodesic conditions, the authors of Sharif \& Siddiqa 
\cite{Sharif} investigated the origins of a gravitational radiation in
Starobinsky model. In this study, we propose to build a cosmological
scenario using a $f(Q,T)$ functional form with the same $Q$ dependency as
the Starobinsky model (with $R$ replaced by $Q$), that is, with a quadratic
additional contribution of $Q$. The T-dependence will be regarded as linear,
with $2\gamma T$, where $\gamma $ is a constant. Because of this, we'll adopt%
\newline
\begin{equation}
f\left( Q,T\right) =Q+\alpha Q^{2}+2\gamma T  \label{f}
\end{equation}

Let's now write the solutions for the material content of our $%
f\left(Q,T\right) $ model, denoted by the symbols $\rho $ and $p$. Using (%
\ref{f}) in (\ref{13}) and (\ref{14}), we obtain

\begin{widetext}
\begin{equation}
\rho =\chi \left[ \gamma \overset{.}{H}-3H^{2}(\gamma -12\alpha \gamma 
\overset{.}{H}+4\pi )-54\alpha (\gamma +4\pi )H^{4}\right]   \label{rho}
\end{equation}

\begin{equation}
p=\chi \left[ (3\gamma +8\pi )\overset{.}{H}\left( 36\alpha H^{2}+1\right)
+3(\gamma +4\pi )H^{2}\left( 18\alpha H^{2}+1\right) \right]   \label{p}
\end{equation}%
\end{widetext}
where $\chi =1/4(\gamma +2\pi )(\gamma +4\pi )$ and $H=\frac{\overset{.}{a}}{%
a}$.

Now, the higher order temporal derivatives of the scale factor must be used
to comprehend the nature of the expansion of the Universe. In cosmology, the
cosmic acceleration is represented in a dimensionless manner by the
deceleration parameter (DP) $q$, which is defined as,%
\begin{equation}
q\left( t\right) =-\frac{1}{aH^{2}}\left( \frac{d^{2}a}{dt^{2}}\right) ,
\label{DP}
\end{equation}%
where the dots are derivatives by cosmic time. If $\overset{..}{a}>0$ (as
recent observations indicate), the expansion of the Universe is considered
to be "accelerating". Therefore, the DP will be negative in this scenario ($%
q<0$).

In addition, the dimensionless expression of the third-order temporal
derivatives of the scale factor is known as the cosmic "jerk parameter", and
it is defined as, 
\begin{equation}
j(t)=\frac{1}{aH^{3}}\left( \frac{d^{3}a}{dt^{3}}\right) .
\label{jerkparameter}
\end{equation}

As the redshift $z$ is a dimensionless variable, it is simple to translate
the time derivatives to the derivatives with respect to $z$ (where $%
1+z=a_{0}/a$, $a_{0}$ being the current value of $a$) for analyzing the
dynamics of the Universe. The expression for the jerk parameter will be
derived from Eq. (\ref{jerkparameter}), 
\begin{equation}
j(z)=1-(1+z)\frac{(h^{2})^{\prime }}{h^{2}}+\frac{1}{2}(1+z)^{2}\frac{%
(h^{2})^{\prime \prime }}{h^{2}},  \label{jerkequation}
\end{equation}%
where $h(z)=\frac{H(z)}{H_{0}}$, $H_{0}$ is the current value of the Hubble
parameter and a prime represents the derivative with respect to $z$.

The parametrization in the current work is carried out with the presumption
that $j$ is a slowly varying amount, and it will be treated as a constant in
the discussion that follows. The formula for $h^{2}(z)$ is given by the
solution of the differential equation (\ref{jerkequation}) as, 
\begin{equation}
h^{2}(z)=A(1+z)^{\frac{3+\sqrt{1+8\,j}}{2}}+B(1+z)^{\frac{3-\sqrt{1+8\,j}}{2}%
}.  \label{h2zABj}
\end{equation}

In Eq. (\ref{h2zABj}), $A$ and $B$ are really the constant dimensionless
parameters. Now, the boundary condition is used to determine the
relationship between $A$ and $B$: $h(z=0)=1$ as $A+B=1$. The final
representation of $h^{2}(z)$ as a function of redshift $z$ and two
parameters $j$ and $A$ is expressed as, 
\begin{equation}
h^{2}(z)=A(1+z)^{\frac{3+\sqrt{1+8\,j}}{2}}+(1-A)(1+z)^{\frac{3-\sqrt{1+8\,j}%
}{2}}.  \label{h2zAj}
\end{equation}

The model parameters are $j$ and $A$, making it basically a two-parameter
model. The value of $j$ determined by the MCMC analysis of the reconstructed
model using various observational data would show the consistency or
departure of this model from the $\Lambda$CDM, and for $j=1$, it perfectly
replicates the $\Lambda$CDM.

The DP, which is described in Eq. (\ref{DP}), may alternatively be
represented for the current model in terms of the redshift and model
parameters as,

\begin{widetext}
\begin{equation}
q(z)=-1+\frac{A\left( \sqrt{8j+1}(z+1)^{\sqrt{8j+1}}+3(z+1)^{\sqrt{8j+1}}+%
\sqrt{8j+1}-3\right) -\sqrt{8j+1}+3}{4A\left( (z+1)^{\sqrt{8j+1}}-1\right) +4%
}.
\end{equation}
\end{widetext} where, we used $q=-1-\frac{\overset{.}{H}}{H^{2}}$ and $\dot{H%
}=-(1+z)H(z)\frac{dH}{dz}$.

\section{Observational data}
\label{sec4}

In this section, we use updated Hubble data sets, Pantheon SNe data sets, and BAO data sets to determine parameter values for our cosmological model. The parameter space $\theta_{s}=(A,j,H_0)$ of our model is explored using the Markov Chain Monte Carlo (MCMC) technique and Bayesian analysis, facilitated by the \textit{emcee} Python package \cite{Mackey/2013}. Furthermore, we incorporate the prior outlined in Tab. \ref{tab} into our analysis. Given that the model parameters $\alpha$ and $\gamma$ are not directly apparent within the Hubble parameter expression provided in Eq. (\ref{h2zAj}), we choose to fix them to specific values in order to investigate the evolution of density, pressure, and the EoS parameter. We adopt the values $\alpha=-0.5$ and $\gamma=-6$, which are consistent with the accelerating Universe scenario.

\subsection{Hubble}
The measurements of Hubble are the initial observational data sample used in
our computation. We are well aware that the Hubble parameter may forecast
the pace of cosmic expansion directly. In general, there are two widely used
methods for calculating the Hubble parameter at given redshifts:
Differential age and the line of sight BAO procedure. In this paper, we
constrain the jerk model using 31 Hubble observations from differential age
procedure in Ref. \cite{Hubble,Sharov}. In the framework of our proposed model, the key model parameters under consideration encompass the coefficients $A$ and $j$, alongside the Hubble constant $H_0$. In light of the complex nature presented by the presence of many free parameters within our model, conducting a comprehensive and comparable observational analysis with previous works imposes additional constraints \cite{Mukherjee1,Mukherjee2,Zhai}. To address this challenge, we make the judicious decision to fix the Hubble constant at a specific value. We align this choice with the most recent Planck data, which provide a strong reference point for cosmological parameters. Specifically, we opt to set $%
H_{0}=(67.4\pm0.5)$ $km/s/Mpc$, a value that is underpinned by the latest Planck measurements and their associated uncertainties \cite{Planck2020}. Several studies in $f(Q,T)$ gravity have adopted a similar methodology to constrain the Hubble constant \cite{Arora3,Koussour1}.

The $\chi ^{2}$ function for Hubble
data points is written as, 
\begin{equation}
\chi _{Hubble}^{2}=\sum\limits_{k=1}^{31}\frac{[H_{th}(z_{k},\theta_{s}
)-H_{obs}(z_{k})]^{2}}{\sigma _{H(z_{k})}^{2}}.  \label{4a}
\end{equation}

In the above~equation, $H_{obs}$ is the Hubble parameter value recovered
from cosmic observations, $H_{th}$ is its theoretical value estimated at $%
z_{k}$ with parameter space $\theta_{s} =(A,j)$, and $\sigma _{H(z_{k})}$ is the
associated error. 

\subsection{SNe}
In our constraints, we also use the newly published Pantheon SNe sets of
data, which has 1048 supernovae samples with distance modulus $\mu ^{obs}$
in the redshift zone $z\in \lbrack 0.01,2.3]$ \cite{Scolnic}. The $\chi^{2}$
function for Pantheon data points is written as, 
\begin{equation}
\chi _{SNe}^{2}=\sum_{i,j=1}^{1048}\Delta \mu _{i}\left(
C_{SNe}^{-1}\right) _{ij}\Delta \mu _{j},  \label{4b}
\end{equation}%
where, $C_{SNe}$ denotes the covariance matrix \cite{Scolnic}, and 
\begin{equation*}
\quad \Delta \mu _{i}=\mu ^{th}(z_{i},\theta )-\mu _{i}^{obs},
\end{equation*}%
is the difference between both the measured distance modulus value obtained
from cosmic data and its theoretical values generated from the model with
the specified parameter space $\theta_{s} =A,j$. The distance modulus is defined
as $\mu =m_{B}-M_{B}$, where $m_{B}$ and $M_{B}$ signify the measured
apparent magnitude and absolute magnitude at a certain red shift (Trying to
retrieve the nuisance parameter using the new BEAMS with Bias Correction
technique (BBC) \cite{BMS}). Its theoretical value is also given by 
\begin{equation}
\mu (z)=5log_{10}\left[ \frac{D_{L}(z)}{1Mpc}\right] +25,  \label{4d}
\end{equation}%
where 
\begin{equation}
D_{L}(z)=c(1+z)\int_{0}^{z}\frac{dz^{^{\prime }}}{H(z^{^{\prime }},\theta_{s} )}.
\label{4e}
\end{equation}%

\subsection{BAO}
BAO observation provided the last constraints in this investigation. BAO
studies oscillations induced in the early Universe by cosmic perturbations
in a fluid composed of photons, baryons, and dark matter that is closely
connected by Thompson scattering. The BAO observations include the Sloan
Digital Sky Survey (SDSS), the Six Degree Field Galaxy Survey (6dFGS), and
the Baryon Oscillation Spectroscopy Survey (BOSS) \cite{BAO1,BAO2}. The
relationships employed in BAO measurements are, 
\begin{equation}
d_{A}(z)=\int_{0}^{z}\frac{dz^{\prime }}{H(z^{\prime })},  \label{4f}
\end{equation}%
\begin{equation}
D_{V}(z)=\left[ \frac{d_{A}(z)^{2}z}{H(z)}\right] ^{1/3},  \label{4g}
\end{equation}%
and 
\begin{equation}
\chi _{BAO}^{2}=X^{T}C_{BAO}^{-1}X  \label{4h}
\end{equation}%
where $C_{BAO}$ symbolizes the covariance matrix \cite{BAO6}, $d_{A}(z)$ the
angular diameter distance, and $D_{V}(z)$ the dilation scale. 

\subsection{Combined analysis}
In our study, we extensively employ various combinations of data sets, particularly the combined Hubble+SNe data sets and the combined Hubble+SNe+BAO datasets. The $\chi^2$ function is formulated for both the Hubble+SNe and the Hubble+SNe+BAO data sets as follows:
\begin{eqnarray}
 \chi _{Hubble}^{2}+\chi _{SNe}^{2},\\
 \chi
_{Hubble}^{2}+\chi _{SNe}^{2}+\chi _{BAO}^{2}.
\end{eqnarray} 
respectively. The model
parameter constraints are derived by minimizing the corresponding $\chi ^{2}$
using MCMC and the emcee library (Tab. \ref{tab} shows the results). During our MCMC analysis, we employed a total of 100 walkers and executed 1000 steps to obtain our results. Fig. %
\ref{Con} shows the $1-\sigma $ and $2-\sigma $ likelihood curves for
the model parameters $A$ and $j$ using Hubble, Hubble+SNe, and
Hubble+SNe+BAO data sets, respectively. The likelihoods are very well
fitted to Gaussian distributions. Moreover, Fig. \ref{Con} and Tab. %
\ref{tab} demonstrate that the best fit value of $j$ is really very near to
one, which indicates that the model with a constant jerk parameter is
temptingly near to the $\Lambda$CDM model. Figs. \ref{ErrorHubble} and \ref%
{ErrorSNe} also exhibit the error bar fitting for the considered model and
the $\Lambda $CDM with $\Omega _{m}^{0}=0.315 \pm0.007$ \cite{Planck2020}. The cosmological transition from a decelerating phase to an accelerating phase is widely attributed to the influence of a cosmic jerk. This critical juncture marks a pivotal moment in the universe's evolution, and its occurrence is closely tied to specific characteristics within different cosmological models. Particularly, this transition is commonly associated with models that exhibit a positive value for the jerk parameter and a concurrent negative value for the deceleration parameter \cite{Chiba,Visser1}.
As a pertinent illustration, consider the $\Lambda $CDM models, which hold a pivotal role in contemporary cosmology. These models are characterized by a constant jerk parameter, specifically $j = 1$. This characteristic configuration implies that the $\Lambda $CDM framework undergoes a seamless transition from a decelerating phase to an accelerating phase, marking a turning point in the cosmic trajectory. Based on these findings, it becomes evident that both datasets, specifically the Hubble data and the combined Hubble+SNe datasets, portray a discernible shift in the jerk parameter value from the foundational $\Lambda $CDM model \cite{Zhai}. However, the constraints gleaned from the dataset that combines Hubble, SNe, and BAO (Hubble+SNe+BAO) reveal a comparatively slight deviation from the $\Lambda $CDM model. This observed discrepancy is of such minute magnitude that it can be confidently regarded as negligible. This discernment leads to a compelling conclusion that the model characterized by a constant jerk parameter gravitates tantalizingly close to the $\Lambda $CDM model. The implications of this proximity are noteworthy, as it suggests that the dynamics of the universe, as manifested through the jerk parameter, maintain remarkable similarity to the well-established $\Lambda $CDM framework. Significantly, these findings draw parallels with the outcomes of a study conducted by Mukherjee and Banerjee \cite{Mukherjee4}, substantiating the robustness and consistency of our results within the broader cosmological discourse.

\begin{widetext}

\begin{figure*}[h]
\subfloat[\label{genworkflow}]{      \includegraphics[trim=1 4 10 20,clip,
width=0.3\textwidth]{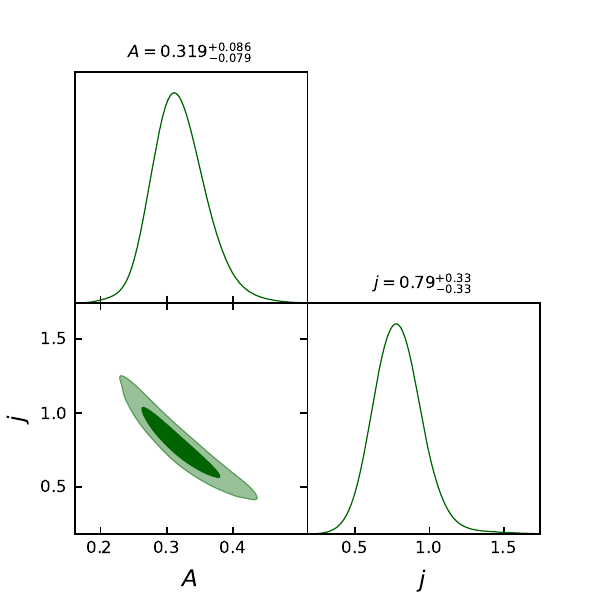}} \hspace{\fill} 
\subfloat[\label{pyramidprocess} ]{      \includegraphics[trim=1 4 10 20,clip,
width=0.3\textwidth]{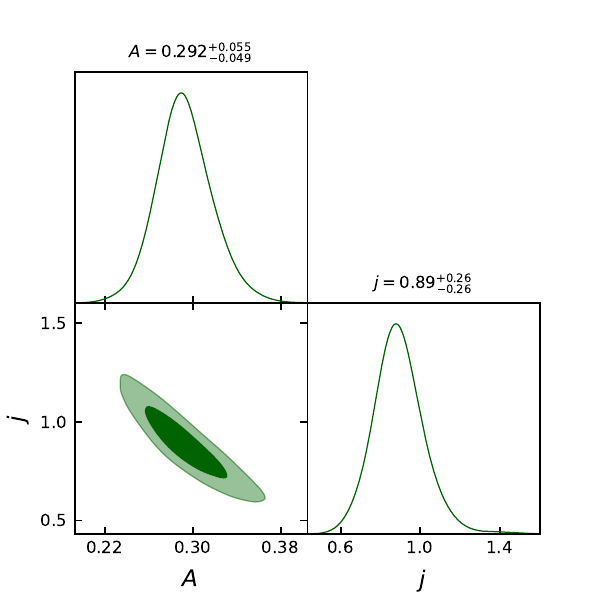}} \hspace{\fill} 
\subfloat[\label{mt-simtask}]{      \includegraphics[trim=1 4 10 20,clip,
width=0.3\textwidth]{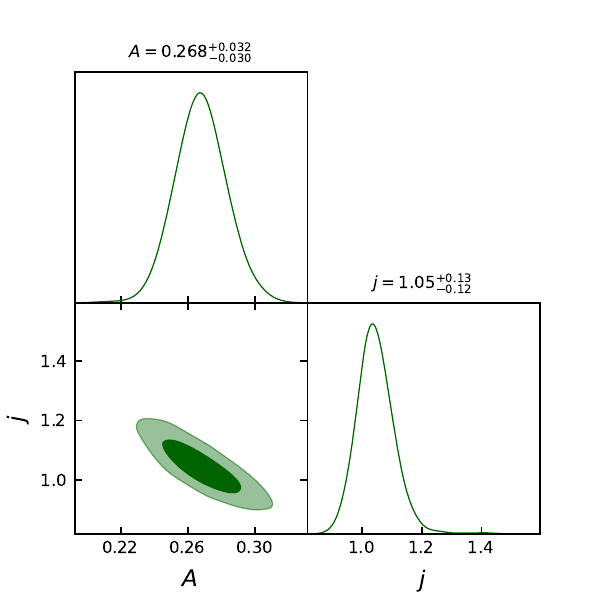}}\newline
\caption{Constraints on the model parameters at $1-\protect\sigma $ and $2-%
\protect\sigma $ confidence interval using the Hubble (a), Hubble+SNe (b), and Hubble+SNe+BAO (c) data sets.}
\label{Con}
\end{figure*}

\begin{table*}[!htbp]
\begin{center}
\begin{tabular}{l c c c c c c}
\hline\hline 
$datasets$              & $A$ & $j$ & $q_{0}$ & $z_{tr}$ & $\omega_{0}$ \\
\hline
$Priors$   & $(0,2)$  & $(0,2)$  & $-$ & $-$ & $-$\\
$Hubble$ & $0.319_{-0.079}^{+0.086}$  & $0.79_{-0.33}^{+0.33}$  & $-0.49^{+0.09}_{-0.08}$ & $0.76^{+0.37}_{-0.36}$ & $-0.92\pm 0.03$\\
$Hubble+SNe$   & $0.292_{-0.049}^{+0.055}$  & $0.89_{-0.26}^{+0.26}$  & $-0.55^{+0.06}_{-0.05}$ & $0.76^{+0.27}_{-0.25}$ & $-0.93\pm 0.01$\\
$Hubble+SNe+BAO$   & $0.268_{-0.030}^{+0.032}$  & $1.05_{-0.12}^{+0.13}$  & $-0.61^{+0.04}_{-0.03}$ & $0.73^{+0.15}_{-0.14}$ & $-0.94\pm 0.01$\\

\hline\hline
\end{tabular}
\caption{The MCMC findings from various datasets have been summarized for a comprehensive overview.}
\label{tab}
\end{center}
\end{table*}

\begin{figure}[h]
\centerline{\includegraphics[scale=0.50]{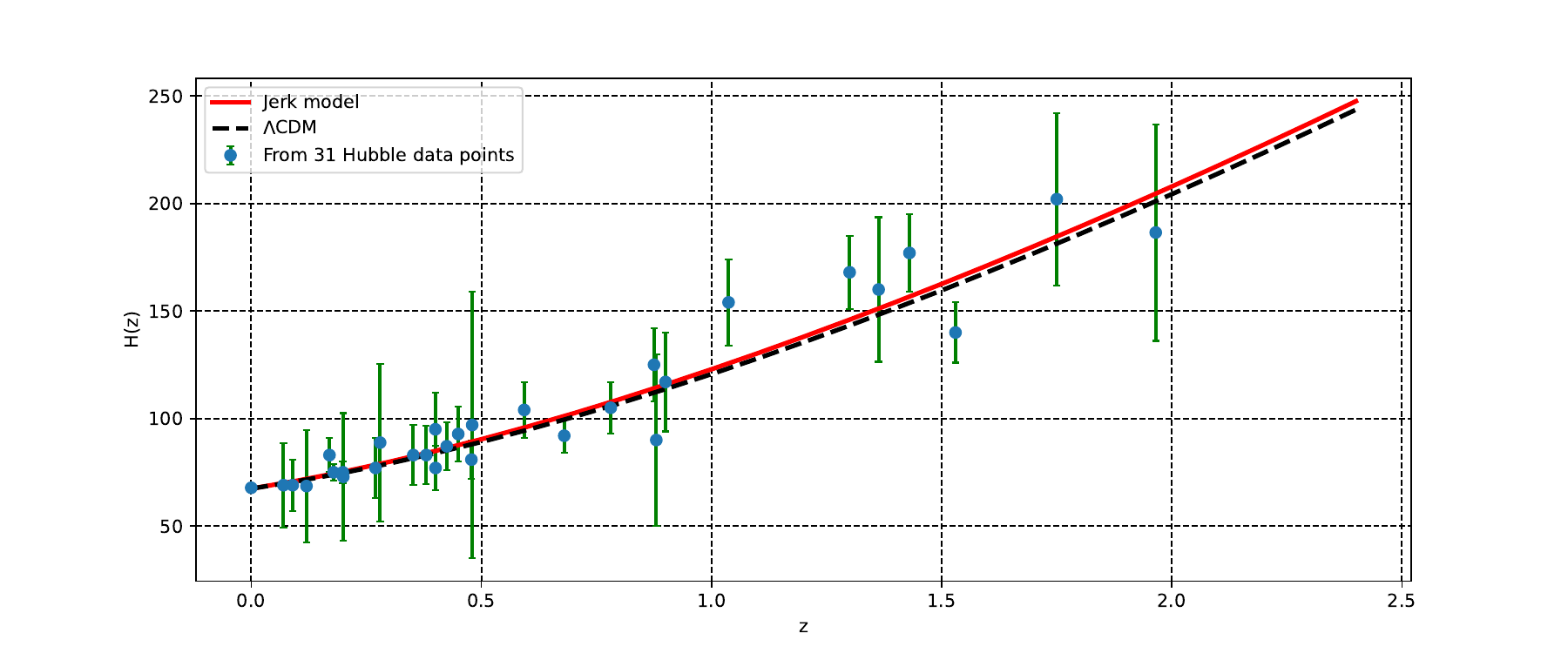}}
\caption{A good fit to the 31 points of the Hubble data sets is displayed in the plot of $H(z)$ versus the redshift $z$ for our $f(Q,T)$ model, which is shown in red, and $\Lambda$CDM, which is shown in black dashed lines.}
\label{ErrorHubble}
\end{figure}

\begin{figure}[h]
\centerline{\includegraphics[scale=0.50]{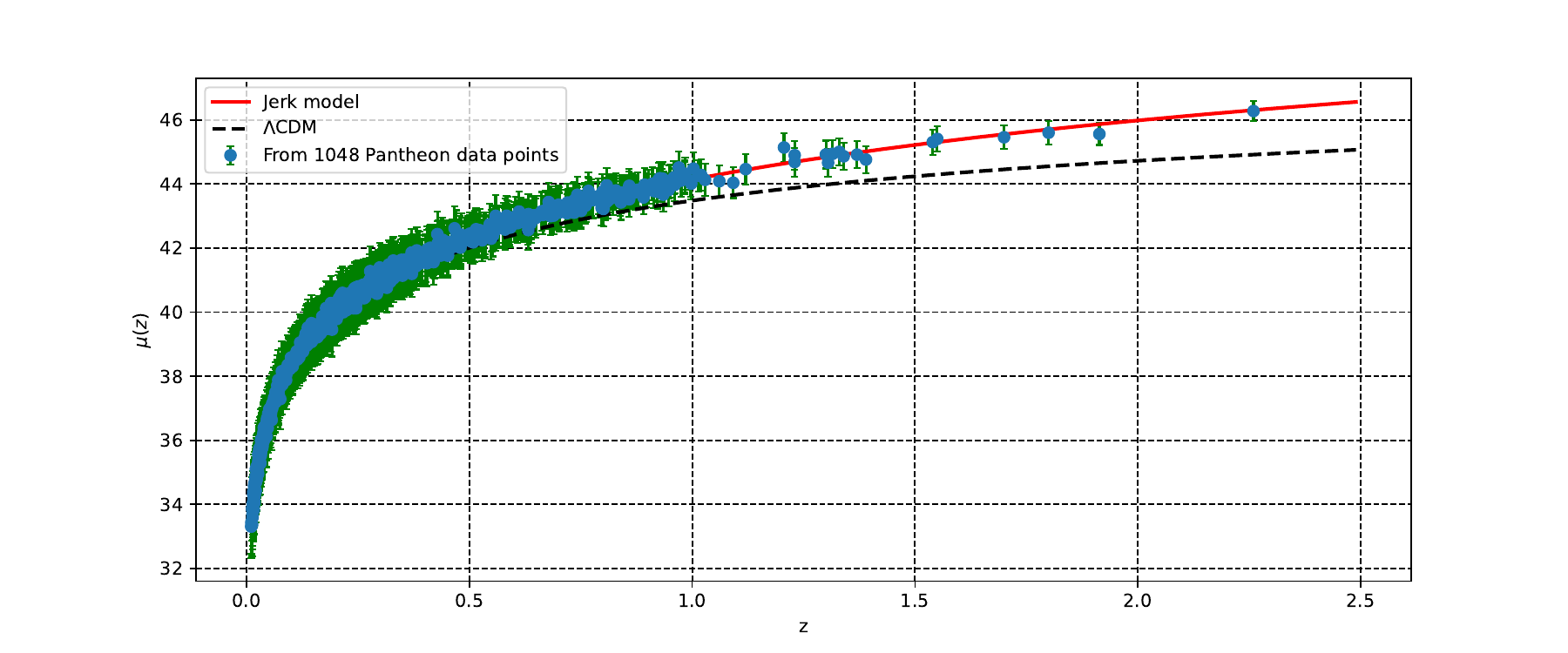}}
\caption{A good fit to the 1048 points of the Pantheon data sets is displayed in the plot of $\mu(z)$ versus the redshift $z$ for our $f(Q,T)$ model, which is shown in red, and $\Lambda$-CDM, which is shown in black dashed lines.}
\label{ErrorSNe}
\end{figure}

\end{widetext}

\begin{figure}[tbp]
\includegraphics[width=8.5 cm]{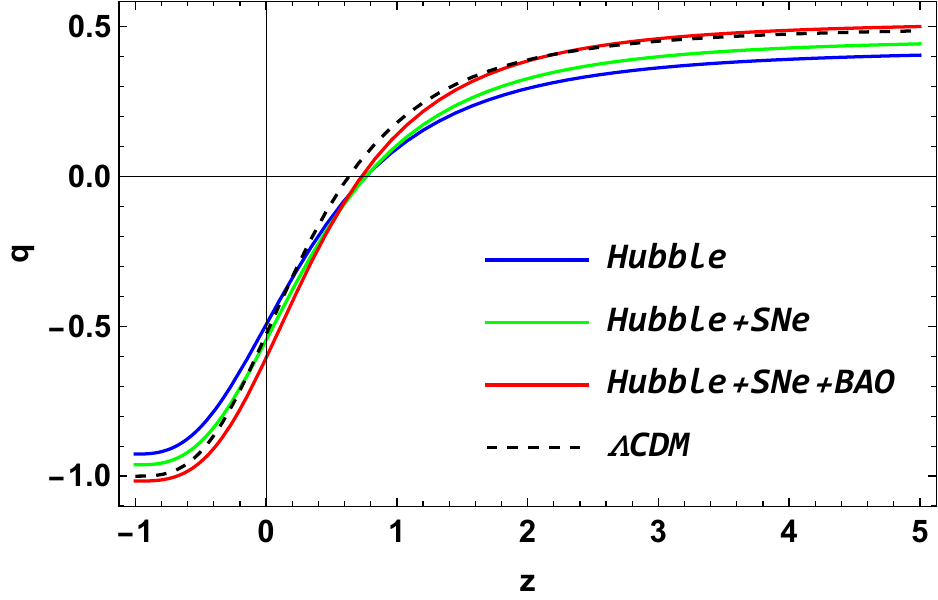}
\caption{The graph above depicts the relationship between the deceleration
parameter ($q$) and redshift ($z$) according to the values of model
parameters constrained by Hubble, Hubble+SNe, and Hubble+SNe+BAO sets of
data.}
\label{FDP}
\end{figure}

\begin{figure}[tbp]
\includegraphics[width=8.5 cm]{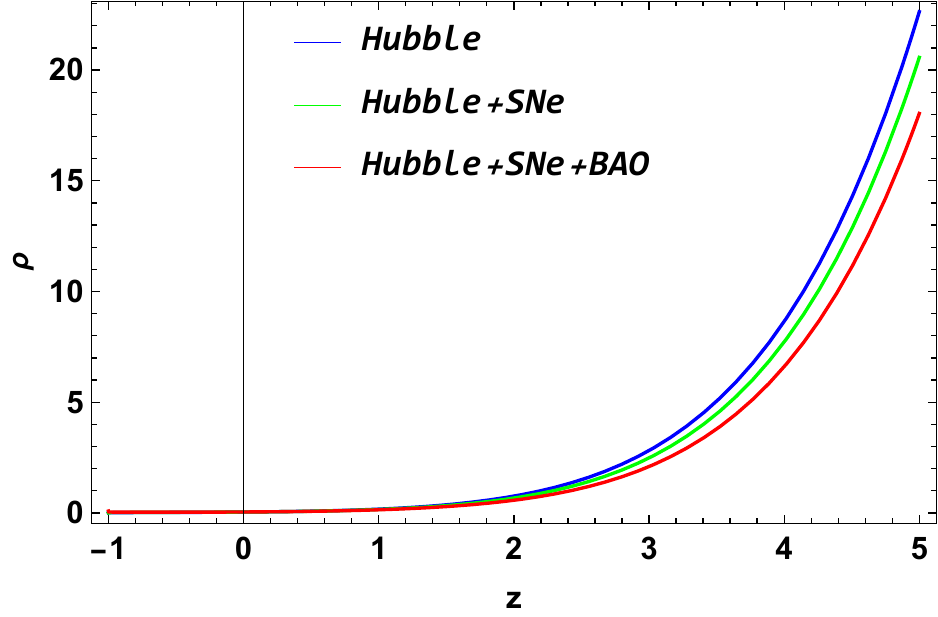}
\caption{The graph above depicts the relationship between the energy density
($\protect\rho $) and redshift ($z$) according to the values of model
parameters constrained by Hubble, Hubble+SNe, and Hubble+SNe+BAO sets of
data.}
\label{Frho}
\end{figure}

\begin{figure}[tbp]
\includegraphics[width=8.5 cm]{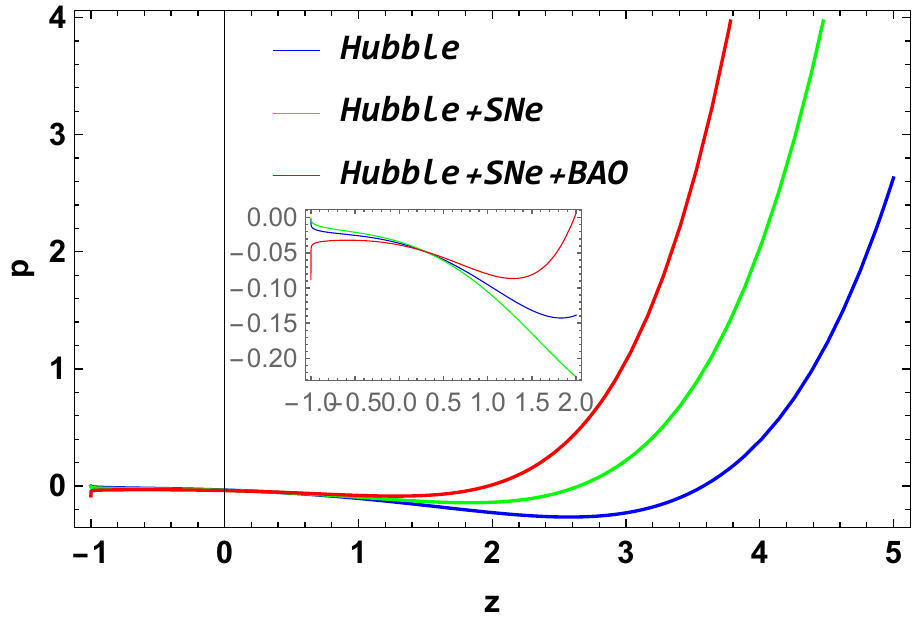}
\caption{The graph above depicts the relationship between the pressure ($p$)
and redshift ($z$) according to the values of model parameters constrained
by Hubble, Hubble+SNe, and Hubble+SNe+BAO data sets.}
\label{Fp}
\end{figure}

The evolution curves of the DP, pressure and energy density according to the
constrained values of the model parameters are shown here. According to Fig. %
\ref{FDP}, the DP is positive in the early Universe and negative in the late
Universe. As a result, it shows that the Universe is transitioning from
deceleration to acceleration. The $q$ increases as cosmic red shift
increases. The transition redshifts associated to the model parameter values
imposed by Hubble, Hubble+SNe, and Hubble+SNe+BAO data sets are estimated
as, $z_{tr}=0.76^{+0.37}_{-0.36}$, $0.76^{+0.27}_{-0.25}$, and $z_{tr}=0.73^{+0.15}_{-0.14}$ respectively.
Furthermore, the current value of the DP is $q_{0}=-0.49^{+0.09}_{-0.08}$ for the Hubble
data sets, $q_{0}=-0.55^{+0.06}_{-0.05}$ for the Hubble+SNe data sets, and $%
q_{0}=-0.61^{+0.04}_{-0.03}$ for the Hubble+SNe+BAO data sets. In addition, it is essential to point out that the $q_{0}$ and $z_{tr}$ values constrained in this paper are consistent with the values reported in Refs. \cite{Capozziello, Mamon, Basilakos}. The energy density
decreases as the Universe expands, as shown in Fig. \ref{Frho}. At late
times ($z\rightarrow -1$), the energy density tends to be zero. Fig. \ref{Fp}
further shows that the pressure of the model is negative in the present ($%
z=0 $) and future ($z<0$). Negative pressure is used to describe the process
of the acceleration of the Universe in modified gravity. In addition, we
discovered that pressure increases with red shift. 

\section{EoS parameter $\protect\omega $ and energy conditions}

\label{sec5}

\begin{figure}[tbp]
\includegraphics[width=8.5 cm]{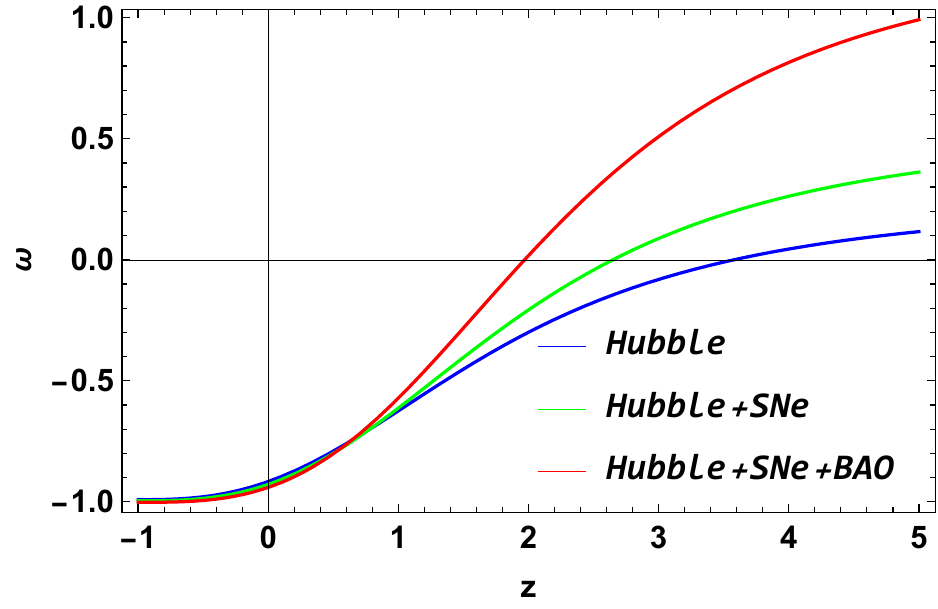}
\caption{The graph above depicts the relationship between the EoS parameter (%
$\protect\omega $) and redshift ($z$) according to the values of model
parameters constrained by Hubble, Hubble+SNe, and Hubble+SNe+BAO sets of
data.}
\label{FEoS}
\end{figure}

The effective or total equation of state (EoS) $\omega$ is determined by dividing the total pressure by the total energy density, i.e. $\omega =\frac{p}{\rho }$. The evolution of the energy density and the expansion of the universe are directly intertwined with the behavior of this parameter. Different EoS values correlate to distinct epochs of the Universe in its
early decelerating and current accelerating expanding stages. It contains
stiff-fluid, radiation, and matter dominated (dust) for $\omega =1$, $\omega
=\frac{1}{3}$, and $\omega =0$ (decelerating stages), respectively. It
depicts quintessence $-1<\omega <-\frac{1}{3}$, the cosmological constant $%
\omega =-1$, and the phantom scenario, $\omega <-1$. By using Eqs. (\ref{rho})
and (\ref{p}) we get the expression for EoS parameter and plot its behavior in
Fig. \ref{FEoS} for the parameter values of the model constrained from three
data sets. Fig. \ref{FEoS} shows that $\omega $ is presently negative and
exhibiting quintessence dark energy, indicating an accelerating phase. It
should be noticed that the EoS parameter in our model tends to $-1$ at late
periods. As a result, it behaves as a cosmological constant at late periods.
Also, the current values of the EoS parameter for the Hubble, Hubble+SNe,
and Hubble+SNe+BAO data sets are $\omega _{0}=-0.92\pm 0.03$, $\omega
_{0}=-0.93\pm 0.01$, and $\omega _{0}=-0.94\pm 0.01$, respectively \cite{Hernandez, Zhang}. 

We know that physical parameters such as DP and EoS parameter are important
in the study of the Universe. Another major research in current cosmology is
on energy conditions derived from Raychaudhuri's equation. The main
objective of these energy conditions is to limit the expansion of the
Universe. Energy conditions include null energy condition (NEC), weak energy
condition (WEC), dominant energy condition (DEC), and strong energy
condition (SEC) (SEC). These energy conditions are described in $f(Q,T)$
modified theory of gravity with specified energy density $\rho $\ and
pressure $p$ as follows: $\rho +p\geq 0$ and $\rho \geq 0$ (WEC); $\rho
+p\geq 0$ (NEC); $\rho \geq \left\vert p\right\vert $ and $\rho \geq 0$
(DEC); $\rho +3p\geq 0$ (SEC). The violation of NEC leads in the violation
of leftover energy conditions (also the violation of the NEC results in the
violation of the second law of thermodynamics), it reflects the depletion of
energy density with the expanding Universe. Further, the violation of SEC
indicates the acceleration of the Universe \cite{Visser}. 
\begin{figure}[tbp]
\includegraphics[width=8.5 cm]{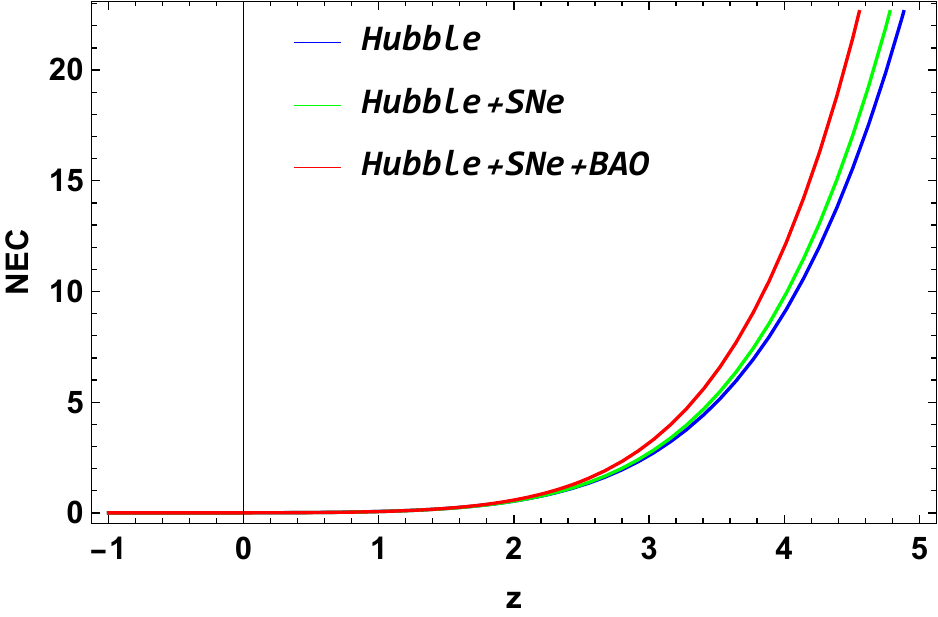}
\caption{The graph above depicts the relationship between the NEC condition
and redshift ($z$) according to the values of model parameters constrained
by Hubble, Hubble+SNe, and Hubble+SNe+BAO data sets.}
\label{FNEC}
\end{figure}
\begin{figure}[tbp]
\includegraphics[width=8.5 cm]{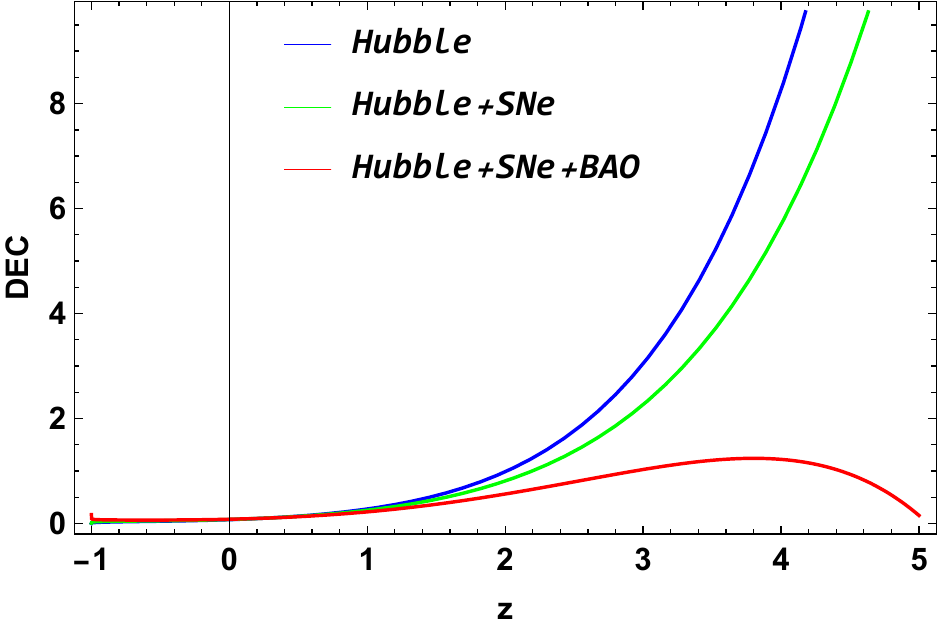}
\caption{The graph above depicts the relationship between the DEC condition
and redshift ($z$) according to the values of model parameters constrained
by Hubble, Hubble+SNe, and Hubble+SNe+BAO data sets.}
\label{FDEC}
\end{figure}
\begin{figure}[tbp]
\includegraphics[width=8.5 cm]{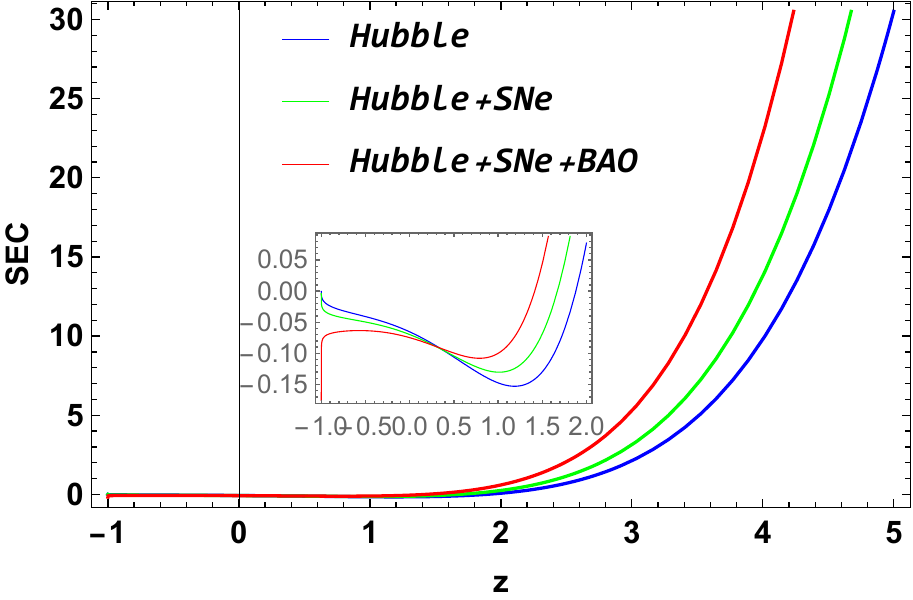}
\caption{The graph above depicts the relationship between the SEC condition
and redshift ($z$) according to the values of model parameters constrained
by Hubble, Hubble+SNe, and Hubble+SNe+BAO data sets.}
\label{FSEC}
\end{figure}

Figs. \ref{FNEC}, \ref{FDEC}, and \ref{FSEC} depict graphs of energy
conditions with regard to redshift, and we can see from these figures that
NEC and DEC are currently satisfied. Also, as seen in Fig. \ref{FSEC}, $\rho
+3p\leq 0$ results in a violation of the SEC at the present. Thus, violating
SEC causes the Universe to accelerate.

\section{$Om(z)$ diagnostic}
\label{sec6}

The $Om(z)$ diagnostic serves as a valuable tool for scrutinizing distinctions between the conventional $\Lambda$CDM model and alternative DE models. This diagnostic method proves to be more user-friendly than the statefinder diagnosis \cite{Sahni}, given its reliance solely on the primary temporal derivative of the cosmic scale factor. This streamlined approach is attributed to its dependence solely on the Hubble parameter, which in turn relies on a solitary time derivative of $a(t)$, the cosmic scale factor. In the context of a spatially flat universe, the definition of the $Om(z)$ diagnostic takes the form:
\begin{equation}
Om\left( z\right) =\frac{h^{2}(z)-1}{\left( 1+z\right) ^{3}-1}.
\end{equation}

\begin{figure}[tbp]
\includegraphics[width=8.5 cm]{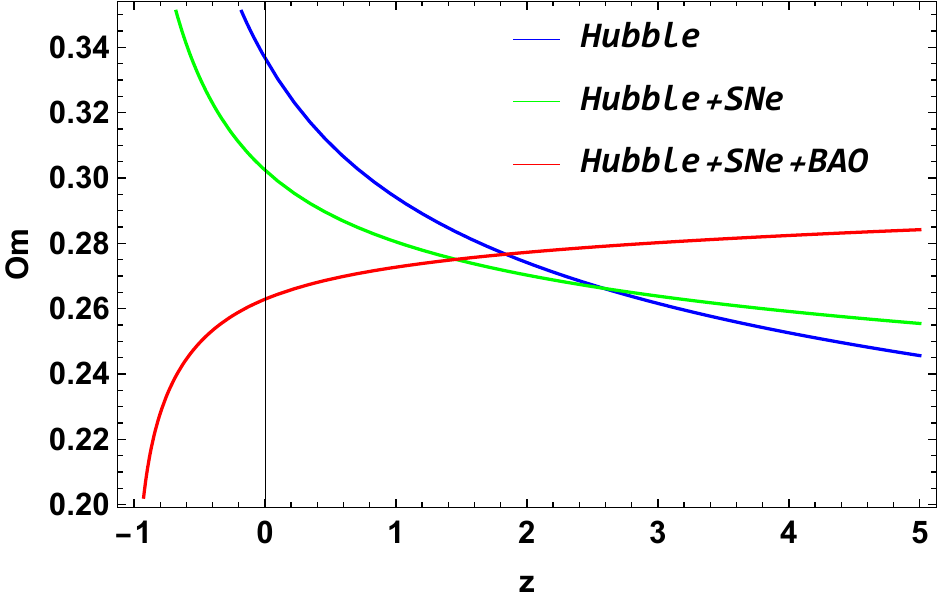}
\caption{The graph above depicts the relationship between the $Om\left(
z\right) $ parameter and redshift ($z$) according to the values of model
parameters constrained by Hubble, Hubble+SNe, and Hubble+SNe+BAO sets of
data.}
\label{FOm}
\end{figure}

Alternatively, we can express that the $Om(z)$ diagnostic provides us with a null test for evaluating the cosmological constant. In the scenario where DE corresponds to a cosmological constant, the behavior of $Om(z)$ becomes linear with a constant slope, specifically $Om(z) = \Omega_{m}^{0}$. However, for alternative DE models, $Om(z)$ takes on a curved trajectory. Specifically, a negative slope within the $Om(z)$ curve indicates quintessence-like behavior, while a positive slope corresponds to a phantom-like behavior. As depicted in Fig. \ref{FOm}, the $Om(z)$ diagnostic, constrained by the limited values of the model parameters extracted from the Hubble and Hubble+SNe datasets, exhibits a consistent negative slope across the cosmic evolution. In this context, the jerk model is indicative of quintessence-like behavior. Conversely, when considering the Hubble+SNe+BAO datasets, the $Om(z)$ diagnostic demonstrates a positive slope, corresponding to a phantom-like behavior of the Universe within this framework.

\section{Concluding remarks}

\label{sec7}

The paper describes the phenomenon of late-time acceleration in the context
of $f(Q,T)$ gravity, which is constructed from an arbitrary function of
non-metricity scalar non-minimally coupled to the trace of the stress-energy
tensor. We have investigated the functional form $f(Q,T)=f\left(Q\right)+f%
\left(T\right)$, where $f\left(Q\right)=Q+\alpha\,Q^2$ is the Starobinsky
model in $f\left(Q\right)$ gravity and $f\left(T\right)=2\,\gamma\,T$, where 
$\alpha$ and $\gamma$ are constants. We considered a constant jerk to get
accurate solutions to the field equations, and then we used the definition
of this jerk to determine the evolution of the other kinematical variables
such as the deceleration parameter, the energy density, EoS parameter, and
different energy conditions to validate the proposed model.

In Sec. \ref{sec4}, to constrain the model parameters $A$ and $j$, we
analyzed the most available Pantheon SNe set of data collection with 1048
points, the Hubble set of data with 31 points, and the BAO set of data with
six points. For the Hubble data sets, the combined Hubble+SNe sets of
data and the combined Hubble+SNe+BAO data sets, we estimated the best fit
values of the model parameters $A$ and $j$. The results obtained are
summarized in Fig. \ref{Con} and Tab. \ref{tab}. Thus, the obtained
data clearly indicate that the best fit value of $j$, which has been chosen
as a constant in this study, is extremely near to one i.e. $j\sim 1$, which
is compatible with a $\Lambda$CDM model. Current findings on constant jerk
parametrization show that a $\Lambda$CDM is clearly recommended \cite%
{jerk1, Mukherjee2}. We also examined the evolution of various cosmological
parameters corresponding to the best fit values of the model parameters. The
deceleration parameter is positive in the early Universe and negative in the
late Universe. Thus, it shows that the Universe is transitioning from
deceleration to acceleration. The transition red shifts associated to the
model parameter values imposed by Hubble, Hubble+SNe, and Hubble+SNe+BAO
data sets are estimated as $z_{tr}=0.76^{+0.37}_{-0.36}$, $z_{tr}=0.76^{+0.27}_{-0.25}$, and $z_{tr}=0.73^{+0.15}_{-0.14}$ respectively. Further, the current values of the DP are $q_{0}=-0.49^{+0.09}_{-0.08}$ for the Hubble
data sets, $q_{0}=-0.55^{+0.06}_{-0.05}$ for the Hubble+SNe data sets, and $%
q_{0}=-0.61^{+0.04}_{-0.03}$ for the Hubble+SNe+BAO sets of
data. The energy density decreases as the Universe expands. So, the negative
pressure is a term used to explain the process of the acceleration of the
Universe in modified gravity.

In Sec. \ref{sec5}, we discussed the EoS parameter, and different energy
conditions to validate the proposed model. It is observed that the EoS
parameter in our model tends to $-1$ ($\Lambda$CDM model) in the late time
evolution. Further, the current values of the EoS parameter for the Hubble,
Hubble+SNe, and Hubble+SNe+BAO data sets are $\omega _{0}=-0.92\pm 0.03$, $\omega
_{0}=-0.93\pm 0.01$, and $\omega _{0}=-0.94\pm 0.01$, respectively. The NEC and
DEC conditions are currently satisfied, while the SEC at the present. Thus,
violating SEC causes the Universe to accelerate.

Finally, in Sec. \ref{sec6}, the Om diagnostics are examined in order to
differentiate between various dark energy theories. It is noted that the $%
Om\left( z\right) $ diagnostic parameter for the Hubble and Hubble+SNe sets
of data has a negative slope throughout of the Universe. Thus, the jerk
model represents quintessence type behavior. Also, for the Hubble+SNe+BAO
data sets the $Om\left( z\right) $ has a positive slope. Thus, it
represents the~phantom type behavior of the Universe. As a consequence, the
research shows that the constant jerk parameter in $f(Q,T)$ Starobinsky
gravity may be employed to generate cosmic acceleration.

\section*{Acknowledgments}
The authors extend their appreciation to the Deputyship for Research \& Innovation, Ministry of Education in Saudi Arabia for funding this research through the project number IFP-IMSIU-2023110. The authors also appreciate the Deanship of Scientific Research at Imam Mohamad Ibn Suad Islamic University (IMSIU) for supporting and supervising this project.

\section*{Data Availability}

All generated data are included in this manuscript.


\end{document}